\documentclass[AMA,STIX1COL]{WileyNJD-v2}
\usepackage{amsmath,amssymb,amsfonts}
\usepackage{graphicx}
\usepackage{textcomp}
\usepackage{enumitem}
\usepackage{xcolor}
\usepackage{times}
\usepackage[binary-units=true]{siunitx}
\usepackage{latexsym}
\usepackage{hyperref}
\hypersetup{colorlinks=true,linkcolor=black,citecolor=blue,filecolor=black,urlcolor=blue}
\usepackage{caption}
\usepackage{subcaption}
\usepackage{tabularx}

\newcommand{\HL}[1]{#1}

\articletype{WORKS19 Special Issue on Workflows in Support of Large-Scale Science GEs:  David Walker, Chase Wu}

\received{}
\revised{}
\accepted{}

\begin{document}

\title{Performance comparison of Dask and Apache Spark on HPC systems \HL{for Neuroimaging}}
	
\author[1]{Mathieu Dugr\'e*}
\author[1]{Val\'erie Hayot-Sasson}
\author[1]{Tristan Glatard}
	
\authormark{Mathieu Dugr\'e \textsc{et al}}
	
\address[1]{\orgdiv{Department of Computer Science and Software Engineering}, \orgname{Concordia University}, \orgaddress{\state{Quebec}, \country{Canada}}}
	
\corres{Mathieu Dugr\'e, Department of Computer Science and Software Engineering, Concordia University, Quebec, Montreal, Canada,\email{mathieu.dugre@concordia.ca}}
	
% \presentaddress{This is sample for present address text this is sample for present address text}
	
\abstract[Summary]{
	The general increase in data size and data sharing motivates the adoption of Big Data strategies in several scientific disciplines.
	However, while several options are available, no particular guidelines exist for selecting a Big Data engine.
	In this paper, we compare the runtime performance of two popular Big Data engines with
	Python APIs, Apache Spark, and Dask, in processing neuroimaging pipelines.
	Our experiments use three synthetic \HL{neuroimaging} applications to process
	the \SI{606}{\gibi\byte} BigBrain image and an actual pipeline to process data
	from thousands of anatomical images.
	We benchmark these applications on a dedicated HPC cluster running the Lustre file system while using varying
	combinations of the number of nodes, file size, and task duration.
	Our results show that although there are slight differences between Dask and Spark,
	the performance of the engines is comparable for data-intensive applications.
	However, Spark requires more memory than Dask, which can lead to 
	slower runtime depending on configuration and infrastructure.
	In general, the limiting factor was the data transfer time.
	While both engines are suitable for neuroimaging, more efforts need to be put
	to reduce the data transfer time and the memory footprint of applications.
}
	
	
\maketitle

This is the accepted version of the following article: Dugr\'e, M., Hayot-Sasson, V., \& Glatard, T. (2023). Performance comparison of Dask and Apache Spark on HPC systems for neuroimaging. Concurrency and Computation: Practice \& Experience, 35(21). \url{https://doi.org/10.1002/cpe.7635}, which has been published in final form at [\url{https://onlinelibrary.wiley.com/doi/10.1002/cpe.7635}]. This article may be used for noncommercial purposes in accordance with the Wiley Self-Archiving Policy [\url{http://www.wileyauthors.com/self}].

	\section{Introduction}
	The rise in data sharing coupled with improved data collection technologies leads neuroimaging into the Big Data era\cite{ALFAROALMAGRO2018400, van2014human, ConpPortal}.
	While standard neuroimaging workflow engines, such as Nipype\cite{Nipype:11}, are well rounded to process standard compute-intensive pipelines,
	they lack Big Data strategies (i.e., in-memory computing, data locality, and lazy-evaluation) to improve the performance of increasingly prevalent data-intensive pipelines.
	A previous study\cite{8752675} noted the importance of in-memory computing and data locality to improve the performance of data-intensive pipelines.
	This work studies how performance benchmarks \HL{of neuroimaging applications} generalize across Big Data engines in an HPC environment.
	\HL{
		With the rising need for Big Data engines in neuroimaging and none commonly
		used in the field, evaluating the performance of different frameworks is essential.
	}
	In particular, we study in-depth the performance difference between Dask\cite{Dask:15} and Apache Spark\cite{Spark:16} for their suitability to process neuroimaging pipelines.
	\HL{
		Our main contribution is to assess whether Spark or Dask has any substantial performance advantage in processing neuroimaging pipelines in HPC systems.
	}
		
	Spark and Dask both provide in-memory computing, data locality, and lazy-evaluation, typical for Big Data engines.
	The scheduler for both engines operates dynamically, which benefits applications with runtime unknown ahead of time\cite{Dask:15}.
	They also provide rich high-level APIs and support a variety of schedulers and deployment environments, such as Mesos\cite{hindman2011mesos}, YARN\cite{vavilapalli2013apache}, Kubernetes, and HPC clusters.
	Although sharing similarities, these engines have differences.
		
	First and foremost, Spark is written in Scala while Dask is in Python.
	The popularity of Python in the scientific community arguably gives an edge to Dask due to the serialization cost between Python and Scala.
	However, if not careful, the Python Global Interpreter Lock (GIL) can significantly reduce the parallelism of an application in some cases.
	The difference in programming languages also provides different benefits.
	On the one hand, as part of the Scipy ecosystem, Dask provides almost transparent integration with APIs such as Numpy arrays, Pandas dataframe, or RAPIDS GPU accelerated code framework.
	On the other hand, Spark's Java, Scala, R, and Python APIs efficiently parallelize pipelines with minimal performance loss due to the Java Virtual Machine.
	Our study is restrained to performance, although we recognize that other factors affect the choice of an engine in practice.
		
	While laptops or workstations can be sufficient for some applications, 
	neuroimaging pipelines often require large\HL{r} infrastructures
	\HL{to process datasets in the scale of several hundreds of {\SI{}{\gibi\byte}} to few {\SI{}{\tebi\byte}}.}
	This paper focuses on the performance of Big Data engines in HPCs environments \HL{for neuroimaging}.
	We installed the Slurm scheduler and Lustre file system~\cite{Braam2019TheLS} on a dedicated cluster to mimic a typical HPC system.
	Lustre is a reference file system for HPC environments, used in over 60 TOP100 supercomputers~{\cite{OpenSFS-lustre}}, offering parallel file accesses.
				
	In a previous study\cite{8943502} we found no significant performance difference between Apache Spark and Dask for data-intensive applications.
	However, using the network file system (NFS) led to an I/O bottleneck limiting application performance.
	Another study~\cite{8588652} also reported network and I/O bottlenecks when concurrency increased.
	This same study found that the startup overhead time for Spark was more prominent than for Dask.
	On the contrary, the work in~\cite{Mehta:17} claims that the startup overhead is more significant for Dask than for Spark.
	This contradiction suggests that the startup overhead might vary depending on cluster configuration.
	The latter study processed a \SI{100}{\giga\byte} dataset using a neuroimaging pipeline and found Dask to be up to 14\% faster than Spark due to ``more efficient pipelining" and serialization time to Python.
	On the contrary, the work in~\cite{10.1145/3225058.3225128} shows that with large datasets, Spark provides better speedup than Dask.
	However, that study used the Dask Bag API, which has performance limitations compared to other Dask APIs\HL{{~\cite{DaskBagDoc}}}.
				
	The above studies compared the performance of the engine at a high level.
	In comparison, our study provides a detailed analysis of the performance differences
	and similarities between Dask and Spark for data-intensive applications executed on HPC systems.
				
	The following section introduces the Lustre file system and the Big Data engines.
	Next, we present the design of our benchmarking experiments.
	We consider two types of neuroimaging pipelines: high-resolution imaging and functional MRI studies for a cohort of subjects.
	The pipelines we chose involve different compute and I/O patterns, e.g., map-reduce or map-only, requiring partial data in memory or the whole dataset.
	Further sections present our results, discussion, and conclusion.

	\section{Background}
	\subsection{Lustre}
	The Lustre file system is a parallel file system of reference in HPC environments.
	Lustre offers a POSIX-compliant interface with the file system.
	Installed in some of the most powerful supercomputers, the design of Lustre allows
	scaling to thousands of nodes, petabytes of storage, and hundreds of gigabits per second of I/O throughput.
					
	At the core of Lustre are the \textit{management servers} (MGSs) and \textit{metadata server} (MDSs) in charge of the file operations and metadata,
	the \textit{object storage targets} (OSTs) store file data, and the \textit{object storage servers} (OSSs) handle the OSTs.
	Lustre clients make requests to the MGSs to access files on the OSTs.
	To handle failures in the system, the MGSs, MDSs, and OSTs have failover capabilities.
	InfiniBand, Ethernet, or a combination of both can be used to interconnect these different components.
					
	While other filesystems such as HDFS or cloud storage could be considered, we chose Lustre to represent an HPC environment best.
	HDFS uses permanent local storage on compute node, which is usually not feasible in HPC.
	Moreover, HPC nodes rarely have access to the internet, limiting cloud storage use.
					
	\subsection{Dask}
	Dask is a Python-based Big Data engine with growing popularity in the scientific Python ecosystem.
	Dask was designed with data locality and in-memory computing in mind to mitigate the data transfer bottleneck in Big Data workflows.
	Data locality, popularized by Map-Reduce~\cite{dean2008mapreduce}, schedules tasks where the data reside.
	In-memory computing minimizes the overhead of transferring data to disks by keeping data in memory.
	Dask uses lazy evaluation to reduce unnecessary communication and computation.
	The engine builds a dynamic graph before execution, allowing it to determine which task to compute.
	Dask workflows can further reduce data transfers by leveraging multithreading whenever the Python GIL does not restrict it.
	It achieves fault tolerance by recording data lineage, the sequence of operations used to modify the initial data.
					
	Dask offers five data structures:
	\href{https://docs.dask.org/en/latest/array.html}{Array},
	\href{https://docs.dask.org/en/latest/bag.html}{Bag},
	\href{https://docs.dask.org/en/latest/dataframe.html}{DataFrame},
	\href{https://docs.dask.org/en/latest/delayed.html}{Delayed},
	and \href{https://docs.dask.org/en/latest/futures.html}{Futures}.
	Arrays offer a clone of NumPy API for distributed processing of large arrays.
	Bags are a distributed collection of Python objects that offers a programming abstraction similar to \href{https://toolz.readthedocs.io/en/latest/}{PyToolz}.
	Dataframes are a parallel composition of \href{https://pandas.pydata.org/}{Pandas} Dataframes used to process a large amount of tabular data.
	Dask Delayed offers an API for distributing arbitrary functions that do not fit the above frameworks.
	Lastly, Dask Futures can also execute arbitrary functions; however, it launches computation immediately rather than lazily.
	Dask allows users to install only required components making it lightweight.
					
	In Dask, a scheduler decides where and when to execute tasks using the Dask graph.
	API operations generate multiple fine-coarse tasks in the computation graph, allowing a more straightforward representation of complex algorithms.
					
	The Dask engine is compatible with multiple distributed schedulers, including YARN and Mesos.
	It also provides its scheduler, \textit{Dask Distributed scheduler}.
	We chose to use Dask Distributed scheduler to keep the environments similar between Dask and Spark.
					
	In the Dask Distributed scheduler, a \textit{dask-scheduler} process administrates the resource provided by \textit{dask-worker}s in the cluster.
	The scheduler receives jobs from clients and assigns tasks to available workers using a global FIFO (First-In-First-Out) job scheduling policy.
	However, the Dask scheduler attempts to complete immediate task dependencies using a LIFO (Last-in-First-Out) policy to reduce the memory footprint of the workers.
					
	Dask offers multiple ways to deploy a cluster, including, but not limited to, SSH configs, Kubernetes, SLURM, PBS.
	For our experiments, we used the \href{https://jobqueue.dask.org/en/latest/generated/dask_jobqueue.SLURMCluster.html}{Dask SLURM cluster} API.
					
	\subsection{Apache Spark}
	Apache Spark is a widely-used general-purpose Big Data engine.
	Like Dask, it aims at reducing data transfer costs by incorporating data locality, in-memory computing, and lazy evaluation.
					
	Spark offers three options to schedule jobs: Spark Standalone, Mesos, and YARN.
	Spark Standalone is a simple built-in scheduler.
	YARN is mainly used to schedule Hadoop-based workflows, while Mesos can be used for various workflows.
	We limit our focus to Spark Standalone scheduler, as researchers are likely to 
	execute their workflows in an HPC environment where neither YARN nor Mesos is usually available.
					
	In the Spark Standalone scheduler, a \textit{leader} (a.k.a master) coordinates the resource provisioned by \textit{workers} in the cluster.
	A \textit{driver}  process receives jobs from clients and requests workers from the leader.
	Jobs are divided into stages to execute onto workers.
	Each operation in a stage is represented by a high-level task in the computation graph.
	Like Dask, Spark Standalone scheduler uses a LIFO policy to schedule tasks.
	Spark Standalone has two execution modes: (1) the client mode, where the driver process runs in a dedicated process,
	and (2) the cluster mode, where the driver runs within a worker process.
	Our experiments use the client mode since cluster mode is not available in PySpark.
					
	Spark's primary data structure is the Resilient Distributed Dataset (RDD)\cite{RDD}, a fault-tolerant, parallel collection of data elements.
	RDDs are the basis of other Spark data structures: Datasets and DataFrames.
	Datasets are similar to RDDs but benefit additional performance by leveraging Spark SQL's optimized execution engine. 
	DataFrames are Datasets organized into named columns and are used to process tabular data. 
	While the DataFrame API is available in all supported languages, Datasets are limited to Scala and Java. 
					
	Python is a standard programming language in the scientific community, offering numerous data processing libraries.
	While serialization from Python to Java, an operation required when using Spark's Python API, creates overhead, we found it minimal in a previous study~\cite{8943502}.
	We focus on PySpark API to have a similar environment between the different engines and  its suitability for neuroimaging.
					
	\section{Methods}
	\subsection{Infrastructure}
	We used the ``slashbin" cluster at Concordia University.
	The cluster has eight compute nodes, four storage nodes, one login node, and one Lustre metadata node.
	Each compute node has two 16-core Intel(R) Xeon(R) Gold 6130 CPU @ 2.10GHz,
	\SI{250}{\gibi\byte} of RAM, \SI{126}{\gibi\byte} of tmpfs,
	% 6 $\times$ SSDs of \SI{447}{\gibi\byte} each with the XFS file system (no RAID enabled),
	CentOS~8.1 and Linux kernel \textit{4.18.0-240.1.1.el8\_lustre.x86\_64}.
	The compute and storage nodes are inter-connected by {\SI{10}{\giga\bit\per\second}} Ethernet link through a Mellanox MSN2410 switch.
	The login node, hosting an NFS server, is connected to the compute nodes through the same switch, with {\SI{10}{\giga\bit\per\second}} Ethernet.
					
	We configured Lustre to have 1 MGS and 4 OSS.
	Each OSS has 11 OSTs (disks) of \SI{8.8}{\tebi\byte}.
	The OSTs are configured to have a maximum of \SI{1000}{\mebi\byte} of dirty data written into the client page cache.
	Both the MDT and OSTs are configured to a maximum of 64 concurrent RPCs in flight.
	A timeout of \SI{100}{\second} is set for Lustre, and OSTs checksum are disabled.
	Lustre \textit{2.14.0-1.el8} is installed on both the Lustre clients and servers.
	We used the default inode configuration for Lustre.
	
	\HL{
		While our infrastructure is not of the scale of TOP500 HPC systems,
		we can assure that no external processes
		impacted our measurements which would be hard to achieve on a production system.
		Our system uses similar hardware as Compute Canada's B\'eluga cluster
		--- a representative cluster available to researchers in Canada --- and 
		replicates as closely as possible its configuration.
		This lets us compare Spark and Dask in an isolated environement with the condition 
		of a realistic  allocation of resources, for neuroimaging applications, in an HPC system.
	}
						
	\HL{
		We configure Dask to have eight worker processes per node, each with eight threads.
		This is the default recommended configuration for the Dask SLURM cluster 
		configuration~{\cite{DaskSLURMDoc}}.
		Similarly, we configured Spark to have eight executors per node
		(hereafter called workers for consistency with Dask), each with eight threads.
		However, when trying different configurations for the workers, we did not
		observe performance differences for our benchmark.
		Therefore we chose the recommended configuration from Dask.
	}
	Each worker was allocated \SI{31.5}{\gibi\byte} of memory.
	For Spark, the JVM heap space uses 10\% of workers' memory, and the location of the log is set on an NFS-mounted directory.
	The worker workspace for both engines is located on Lustre due to the limited fast storage space on the worker nodes.
	Other Dask configurations are left to default.
	A new Dask or Spark cluster was started and torn down for each experiment.
	Dask \textit{2021.7.0} and Spark \textit{3.1.2} were used.
							
	\subsection{Dataset}
	We used BigBrain\cite{Amunts:13}, a 3-D image of the human brain with voxel
	intensities ranging from 0 to 65,535. We converted the blocks into the
	NIfTI format, a popular format in neuroimaging. We left the NIfTI blocks
	uncompressed, resulting in a total data size of \SI{606}{\gibi\byte}. To
	evaluate the effect of block size, we split the image into 1000, 2500,
	and 5000 files of \SI{606}{\mebi\byte}, \SI{242.4}{\mebi\byte}, and
	\SI{121.2}{\mebi\byte}, respectively.
	We used the \href{https://github.com/big-data-lab-team/sam}{sam}~{\cite{sam}} library to split the image.
								
	We also used the dataset provided by the Consortium for Reliability and
	Reproducibility
	(\href{http://fcon_1000.projects.nitrc.org/indi/CoRR/html/}{CoRR})
	\cite{zuo2014open}, freely available on
	\href{https://datasets.datalad.org/?dir=/corr/RawDataBIDS}{DataLad}~{\cite{DataladDataset}}.
	The entire dataset is \SI{379.83}{\gibi\byte}, containing anatomical, diffusion,
	and functional images of 1,654 subjects acquired in 35 sites.
	We used all 3,491 anatomical images, representing \SI{26.67}{\gibi\byte} overall
	(\SI{7.82}{\mebi\byte} per image on average).
	
	\HL{
		We chose these dataset as they represent common dataset sizes for neuroimaging 
		imaging studies and are open-access.
	}
								
	\subsection{Applications}
	This section describes the applications used to benchmark Dask and Spark.
	We use three simple synthetic applications to have a deep understanding of the 
	underlying behavior of the applications with simple I/O patterns, namely
	\textit{Increment}, \textit{Multi-Increment}, and \textit{Histogram}.
	Also, two neuroimaging applications are used to study more realistic 
	applications, namely \textit{Kmeans} and \textit{BIDS App Example}{\cite{gorgolewski2017bids}}.
	\HL{
		We chose these neuroimaging data processing applications to study the 
		effects of various Big Data characteristics.
		Table~{\ref{table:applications}} depicts the Big Data characteristics used in
		each application. 
		In this table, \textit{Map} and \textit{Reduce} represent the Map-Reduce paradigm operations,
		and the \textit{Shuffle} column represents the amount of data shuffled between stages if any.
		The column \textit{Dataset in-memory} depicts whether the whole dataset is required to 
		be in memory for processing.
		Then \textit{Compute intensive} and \textit{Data intensive} shows if the applications are
		compute or data intensive.
		Lastly, the \textit{Containerized} column shows applications wrapped in a container.
	}
	
	\begin{table*}[t]
		\caption{Big Data characteristics of the applications}
		\centering
		\begin{tabular}{l|c|c|c|c|c|c|c|}
			\cline{2-8}
			\multicolumn{1}{c|}{} &
			\textbf{Map} &
			\textbf{Reduce} &
			\textbf{\begin{tabular}[c]{@{}c@{}}Data\\ shuffling\end{tabular}} &
			\textbf{\begin{tabular}[c]{@{}c@{}}Dataset \\ in-memory\end{tabular}} &
			\textbf{\begin{tabular}[c]{@{}c@{}}Compute\\ intensive\end{tabular}} &
			\textbf{\begin{tabular}[c]{@{}c@{}}Data\\ intensive\end{tabular}} &
			\textbf{Containerized} \\ \hline
			\multicolumn{1}{|l|}{\textbf{Increment}}        & X &   &         & Partial &   & X &   \\ \hline
			\multicolumn{1}{|l|}{\textbf{Histogram}}        & X & X & Minimal & Partial &   & X &   \\ \hline
			\multicolumn{1}{|l|}{\textbf{Kmeans}}           & X &   & X       & Whole   & X &   &   \\ \hline
			\multicolumn{1}{|l|}{\textbf{BIDS App Example}} & X & X & Minimal & Partial & X &   & X \\ \hline
		\end{tabular}
		\label{table:applications}
	\end{table*}
				
	\HL{
		To minimize the performance difference from implementation decisions, the
		code base of the applications is common for both engines.
		Therefore, the performance observed comes directly from the engine scheduling and
		their APIs.
	} The scripts used for our experiments and the results generated are available on
	GitHub at: \mbox{\href{https://github.com/mathdugre/paper-big-data-engines}{https://github.com/mathdugre/paper-big-data-engines/}}.
							
	\subsubsection{Increment}
	We adapted the increment application used in~\cite{8752675}.
	This synthetic application reads blocks of the BigBrain from Lustre and
	simulates computation by sleeping for a specified period (Fig.~\ref{fig:graph-increment}). To simulate
	intermediate results, we repeat a {\SI{1}{\second} sleep delay during each incrementation}.
	We prevent data caching of the blocks by incrementing their voxels
	value by one after each sleep operation. Finally, we write the resulting
	NIfTI image back to Lustre. This application allows us to study the engines
	when their inputs are processed independently. The map-only pattern
	mimics the processing of multiple independent subjects in parallel.
												
	\subsubsection{Histogram}
\begin{figure}[h]
		\centering
		\begin{minipage}{0.4\textwidth}
			\centering
			\includegraphics[width=\linewidth]{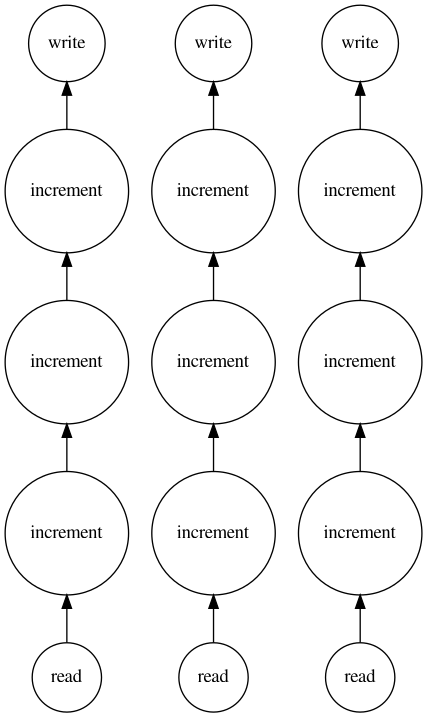}
			\caption{Task graph for Incrementation with 3 iterations and 3 BigBrain blocks.}
			\label{fig:graph-increment}		
		\end{minipage}
		\begin{minipage}{0.55\textwidth}
			\centering
			\includegraphics[width=\linewidth]{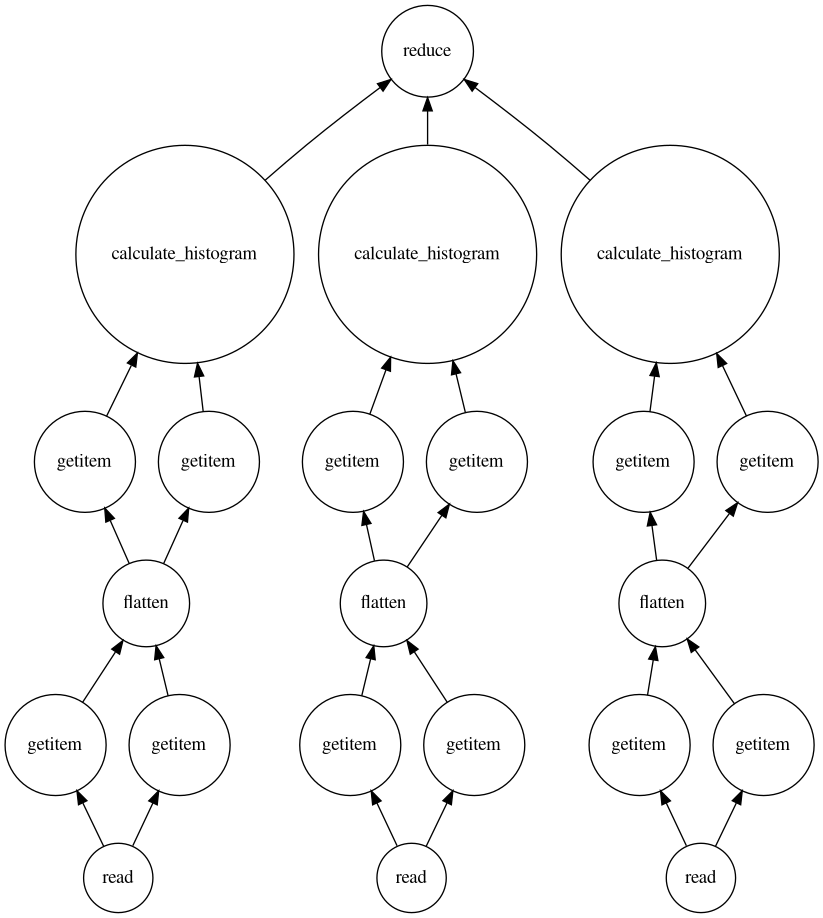}
			\caption{Task graph for Histogram with 3 BigBrain blocks.}
			\label{fig:graph-histogram}			
		\end{minipage}
	\end{figure}

	As our second application, we calculate the histogram of the BigBrain image. 
	The application reads the BigBrain blocks from Lustre, calculates each intensity's frequency, and then writes the aggregated result back on Lustre (Fig.~\ref{fig:graph-histogram}).
	This map-reduce application has a very high read over write ratio.
	Moreover, this application requires shuffling, albeit of a limited amount of data. 
	The amount of inter-worker communication is in-between the increment and multi-increment applications.
									
	\subsubsection{Kmeans}
	For our third application, we apply Kmeans clustering to the voxel
	intensities of the BigBrain image. We set the number of clusters to 3. The
	application starts by reading the image blocks, combining all voxels in a
	1-D array, and choosing initial centroids using the min, max, and
	intermediate values. It assigns each voxel to its closest centroid and
	updates each centroid with the average of its assigned voxels. It repeats
	the assignment and updates steps for a configurable number of iterations.
	Finally, the voxels of the image blocks are classified and written back to
	the file system. Updating the centroids involves substantial data
	communication between the workers.
	The computation to find the closest centroids for each voxels scales with the number of centroids.
	It can be computed in parallel at the cost of higher memory or sequentially to reduce memory usage.
	We decided to compute the nearest centroids in parallel since we use only three centroids.
	However, due to resources limitation on our infrastructure, the Kmeans experiments
	are performed with only half of the BigBrain dataset ({\SI{303}{\gibi\byte}}).
	\HL{We chose this application as it involves substantial data shuffling between workers,
	and requires to keep all the data in memory.}
									
	\subsubsection{BIDS App Example}
	Our fourth application is the BIDS App Example, a neuroimaging pipeline to
	measure brain volume from MRIs. For this application, we use the CoRR
	dataset. The application extracts the brain volume of each participant
	then computes the average for each group of participants. Unlike the other
	applications, the BIDS App Example is a command-line executed in a Docker image
	(bids/example on DockerHub). We converted the Docker image to a Singularity
	image for use in HPC environments, using
	\href{https://hub.docker.com/r/singularityware/docker2singularity/tags/}{docker2singularity}
	\HL{We chose this application for its real-life use case and the fact that it is containerized unlike our other applications.}
											
	\subsection{Experiments}
	Table~\ref{table:parameters} shows the parameters that we varied
	throughout the experiments. We varied (1) the number of nodes to assess
	the scalability of the engines' scheduler, (2) the number of files \HL{to segment}
	BigBrain to measure the effect of different I/O patterns and parallelization degrees, and (3) the
	number of iterations to evaluate the effect of the number of tasks.
	It should be noted that increasing the number of iterations for a given sleep 
	delay also increases the total compute time.
	\HL{
		The choice of parameters for our experiments was restrained by the datasets
		and resources available from the cluster.
		The unrestrained parameters were chosen to have an expected linear scaling
		to facilitate the interpretation of the results.
	}
													
	To avoid potential external biases due to caching, background processes and,
	network load, we ran the applications in randomized order and cleared the
	page cache between every experiment.
	Moreover, we ran each benchmark ten times.
													
	For each execution, we measured the time spent in the different functions of the
	application to read, process, and write data.
	We calculate the overhead for each CPU thread by subtracting the total time
	spent in the reading, writing, and processing functions to the makespan of the application.
	Summing those results gives the total overhead for the application.
													
	\begin{table*}[t]
		\renewcommand{\arraystretch}{1.5}
		\caption{Parameters for the experiments}\label{table:parameters}
		\centering
		\begin{tabular}{|l|c|c|c|c|c|}
			\hline           & Increment & Histogram & Kmeans & BIDS App Example \\\hline
			Dataset size [\SI{}{\gibi\byte}] &\multicolumn{2}{c|}{606} & 303 & \multicolumn{1}{c|}{26.67} \\\hline
			\# of Nodes & \multicolumn{4}{c|}{2, 4, 8} \\\hline
			Number of file & \multicolumn{2}{c|}{1000, 2500, 5000} & 5000 & \multicolumn{1}{c|}{3491} \\\hline
			\# of Iterations & 1, 5, 25  & -         & 1, 3   & -                \\\hline
		\end{tabular}
	\end{table*}
													
	\section{Results}
	\subsection{Increment}
\begin{figure}[!h]
		\centering
		\includegraphics[clip,width=0.75\columnwidth]{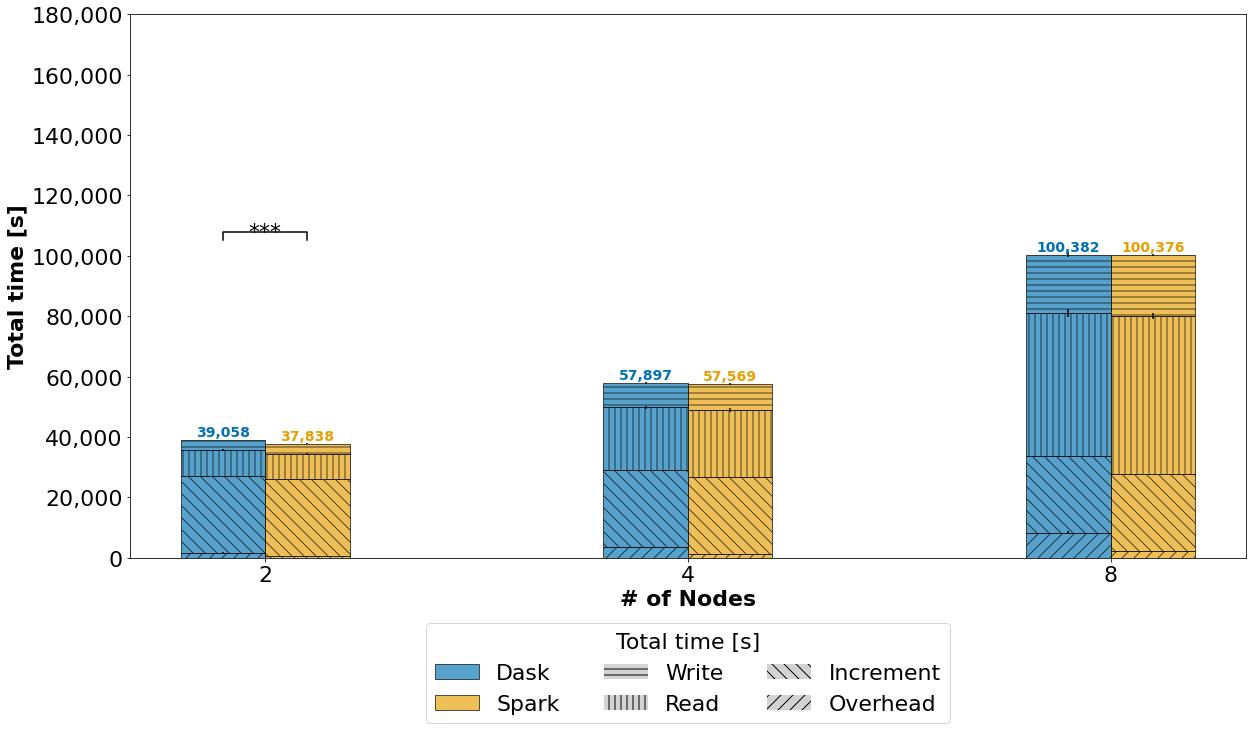}
		\caption{Increment: varying nodes -- 5000 files, 5 iterations}
		\label{fig:increment_worker}
	\end{figure}

	Figure \ref{fig:increment_worker} depicts the total execution time spent in each code segment of the \textit{Increment} application.
	The bars are broken down into Overhead, Read, Write, and other application-specific functions.
	The mean of each bar is calculated from ten repetitions, with the error bar representing the standard deviation.
	As expected, the compute time, from \textit{Increment}, remains the same when we vary the number of nodes.
	However, the \textit{Read} and \textit{Write} time increases with the number of nodes for both engines.
	\HL{
		Due to the large amount of worker threads accessing the Lustre file system
		concurrently, the disk bandiwdth or network are saturated, which results in an I/O bottleneck.
	}
													
	From Figure \ref{fig:increment_worker}, we also observe that Dask's overhead is higher than Spark's.
	While the overhead of both engines increases with the number of nodes, Dask's overhead is significantly worse when scaling.
	Although Dask has a higher overhead, the total time for both engines is overall similar.
	This is due to a tradeoff between disk bandwidth and compute/overhead time.
	\HL{
		On the one hand, when multiple threads access Lustre concurrently they may saturate the shared disk bandwidth, resulting in longer I/O time.
		On the other hand, when most threads are either idle or performing computations,
		the contention on the disk bandwidth is reduced, resulting in shorter I/O time.
	}
	The only significant difference in total time is when two nodes are used.
	This results in an approximately 3\% total runtime difference.
															
\begin{figure}[!h]
		\centering
		\includegraphics[clip,width=0.75\columnwidth]{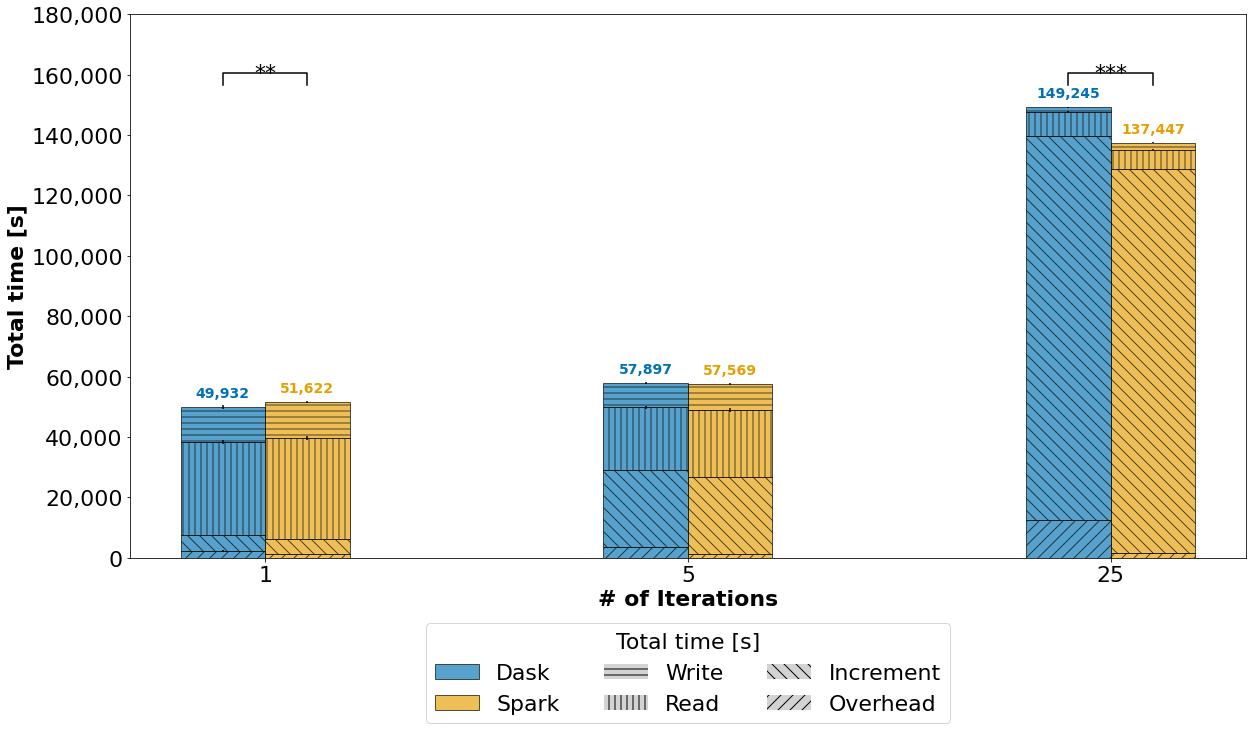}
		\caption{Increment: varying iterations -- 4 nodes, 5000 files}
		\label{fig:increment_itr}
	\end{figure}

	Figure \ref{fig:increment_itr} shows the detailed execution time for the \textit{Increment} application while varying the number of iterations.
	As expected, the overhead for both engines increases with the number of iterations.
	However, overhead for Dask increases significantly faster than Spark, resulting in a 7\% makespan difference for 25 iterations.
															
	Figure \ref{fig:increment_itr} also shows that the total I/O time decreases as total computation time increases.
	This is explained by an I/O bottleneck on the Lustre file system.
	When threads are busy with computations, it reduces the contention on Lustre, thus improving I/O speed for individual accesses.
															
\begin{figure}[!h]
		\centering
		\includegraphics[clip,width=0.75\columnwidth]{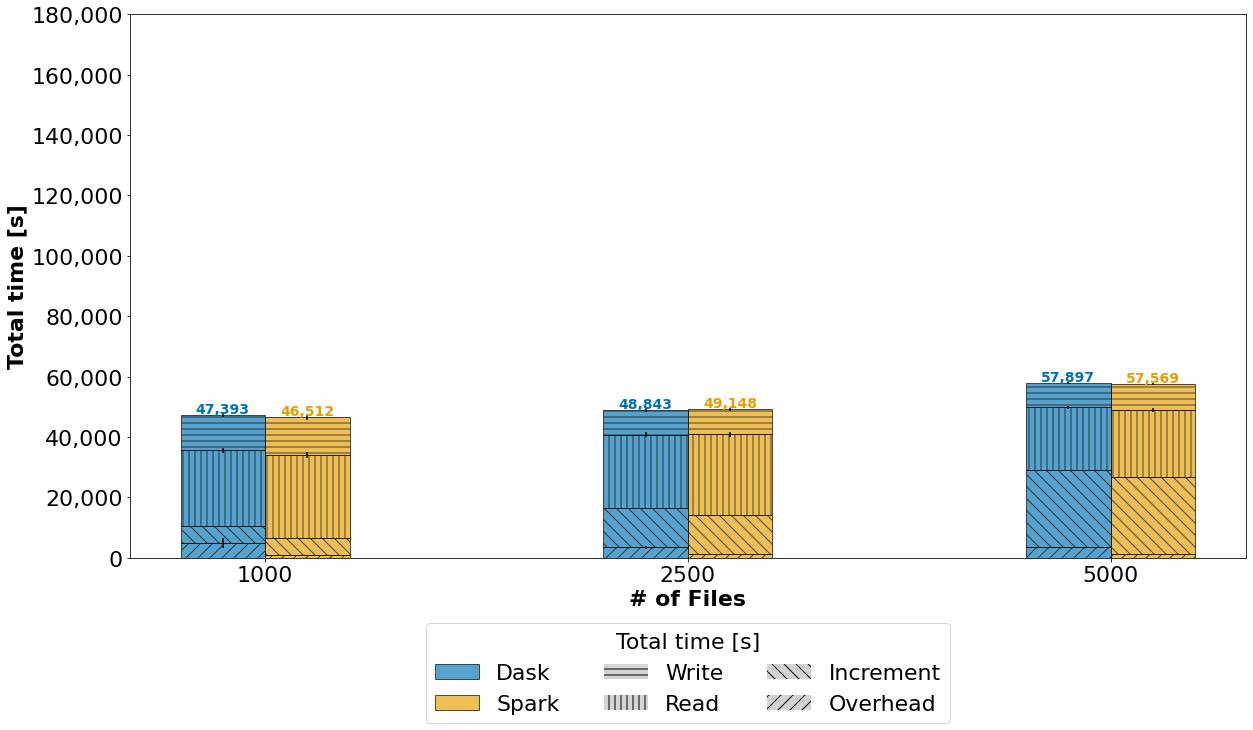}
		\caption{Increment: varying \# of files -- 4 nodes, 5 iterations}
		\label{fig:increment_block}
	\end{figure}

	Figure \ref{fig:increment_block} shows an increase in \textit{Increment} time.
	This is because the application performs an incrementation for each file, thus increasing the computation time.
	Similarly to Figure \ref{fig:increment_itr}, increasing the total compute time of the application reduces the total I/O time.
	Again, this is explained by the I/O tradeoff between compute time and contention of disk bandwidth.
													
	From Figure \ref{fig:increment_block} we also observe that Dask has significantly higher overhead than Spark.
	However, the total execution time of the application is equivalent due to the tradeoff between I/O and compute/overhead time.
																	
	\subsection{Histogram}
\begin{figure}[!h]
		\centering
		\includegraphics[clip,width=0.75\columnwidth]{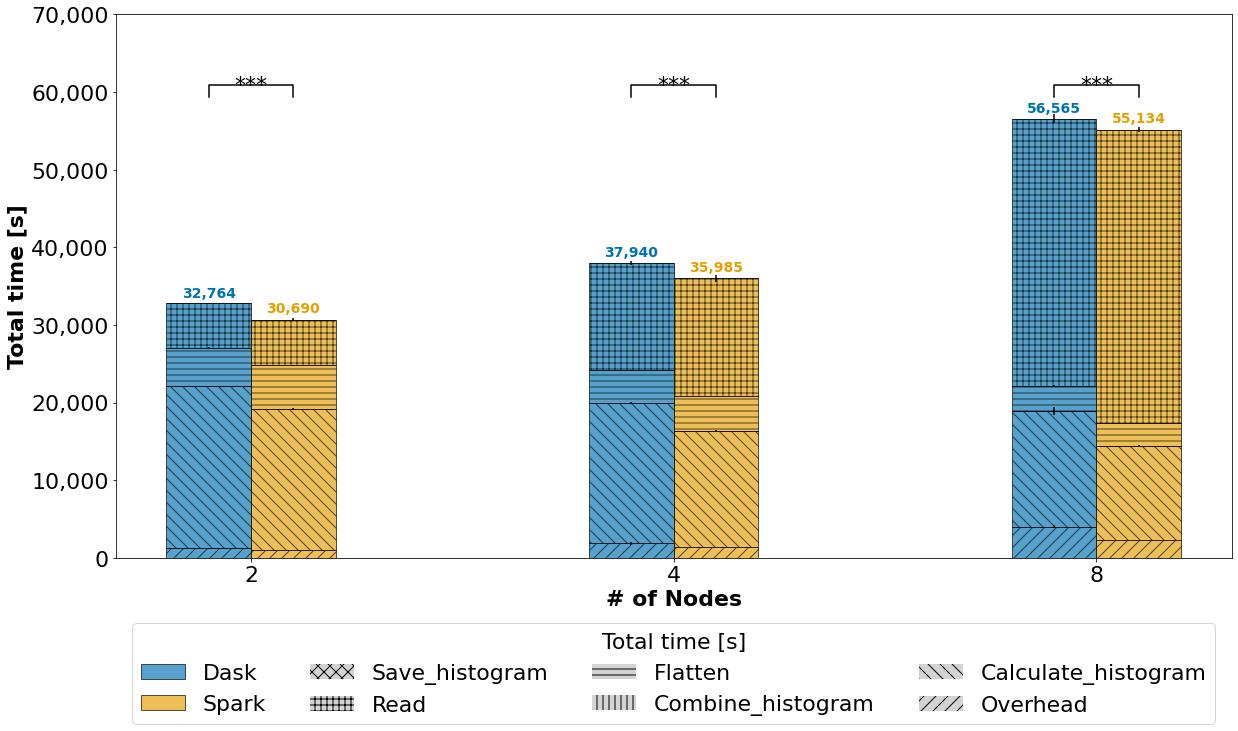}
		\caption{Histogram: varying nodes -- 5000 files}
		\label{fig:histogram_worker}
	\end{figure}

	Figure \ref{fig:histogram_worker} depicts the execution time for the \textit{Histogram} application when varying the number of nodes.
	Overall, the execution time for Spark is significantly lower than for Dask by up to 5\%, with the difference coming from a shorter execution time for \textit{Calculate\_histogram}.
	This is unexpected as the code base for both engines is the same, with only the calling APIs differing.
	It is most likely due to the use of Dask Bags, which are slower than the other Dask APIs\HL{~{\cite{DaskBagDoc}}}.
	As with previous experiments, the overhead for Spark is less and scales better, although compensated by a longer total I/O time.
																
\begin{figure}[!h]
		\centering
		\includegraphics[clip,width=0.75\columnwidth]{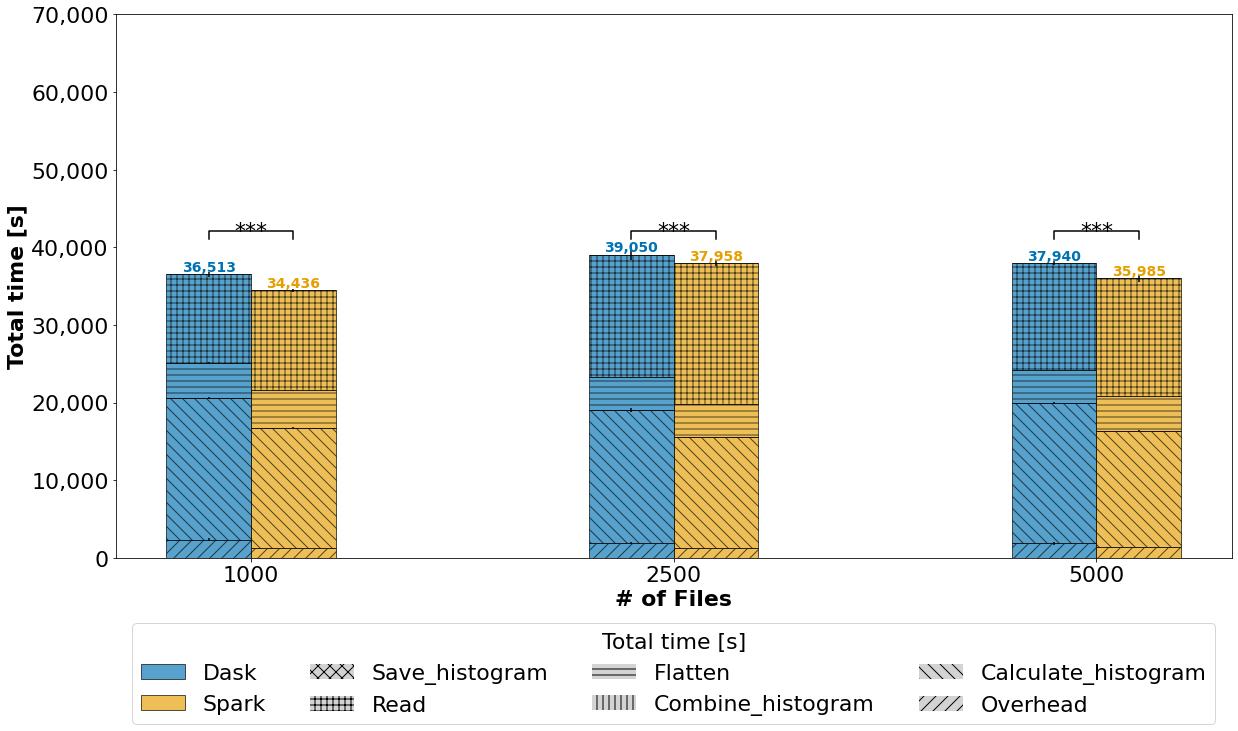}
		\caption{Histogram: varying \# of files -- 4 nodes}
		\label{fig:histogram_block}
	\end{figure}

	Figure \ref{fig:histogram_block} shows the execution time for the \textit{Histogram} when varying the number of files.
	The results are similar to varying the number of nodes with Spark having faster compute time for \textit{Calculate\_histogram}.
	This results in a significant makespan difference with Spark being up to 3\% faster.
														
	\subsection{Kmeans}
	\HL{
		When varying the number of workers, we observed that Dask is multifold faster
		than Spark, resulting in a $1.79\times$ to $2.71\times$ speed-up ratio on the makespan.
		The problem we observed was twofold.
		Firstly, the process ID of the workers' threads is not reported correctly and 
		leading to falsely excessive overhead for Spark.
		Secondly, the Spark workers were restarting due to a lack of memory, leading 
		to task recomputation and having to reload data into memory.
	}
								
	We increase the worker memory by reducing the number of worker processes per
	compute node since all available memory from the compute node is already allocated to the pool of workers.
	When reducing the number of processes from 64 to 32 for Spark and Dask 
	(effectively doubling the memory of each process), the difference in execution time between Spark and Dask was reduced significantly.
	With only 32 processes and more memory per worker, Spark no longer recomputed tasks for the Kmeans application.
	We suppose that given more memory per thread/process, Dask and Spark would then perform similarly.
								
\begin{figure}[!h]
		\centering
		\href{https://mathdugre.me/paper-big-data-engines/ccpe/dask_kmeans_gantt.html}
		{\includegraphics[clip,width=\columnwidth]{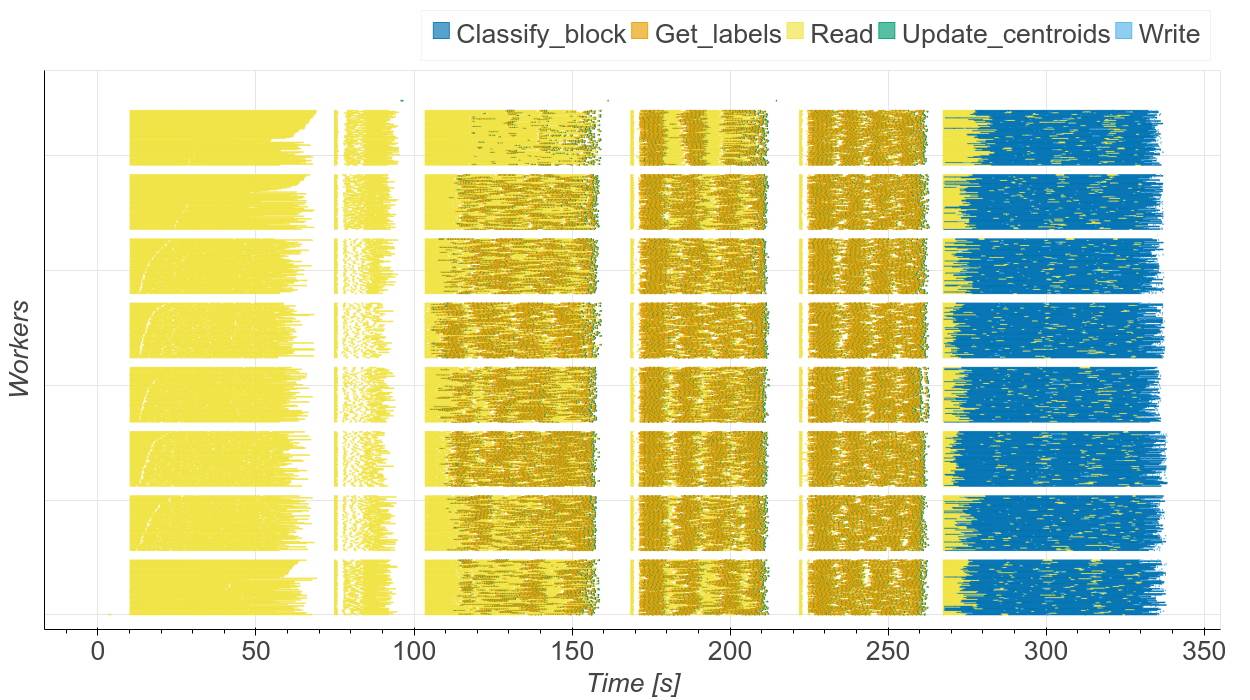}}
		\caption{Kmeans: Dask with 8 nodes, 5000 files, and 3 iterations}
		\label{fig:dask_kmeans_gantt}
	\end{figure}

\begin{figure}[!h]
		\centering
		\href{https://mathdugre.me/paper-big-data-engines/ccpe/spark_kmeans_gantt.html}
		{\includegraphics[clip,width=\columnwidth]{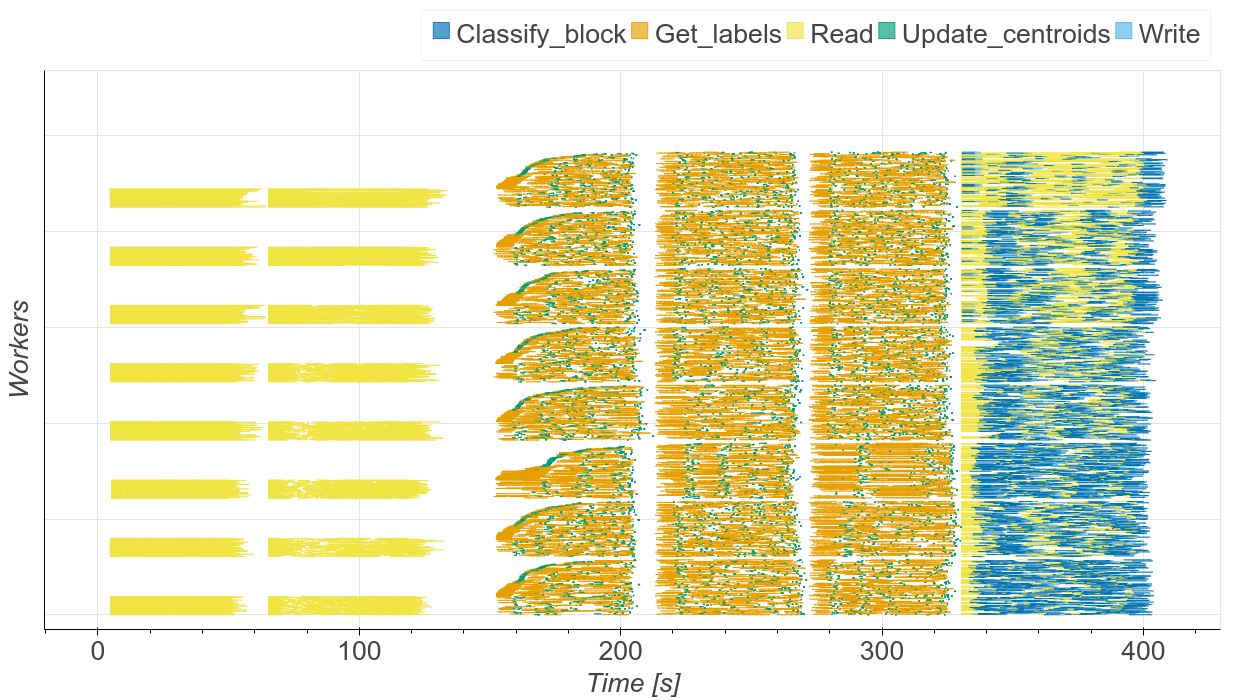}}
		\caption{Kmeans: Spark with 8 nodes, 5000 files, and 3 iterations}
		\label{fig:spark_kmeans_gantt}
	\end{figure}

	In a follow-up experiment, we further optimized our Kmeans implementations for both Dask and Spark to minimize the size of data kept in memory.
	Figure~{\ref{fig:dask_kmeans_gantt}} and Figure~{\ref{fig:spark_kmeans_gantt}} show the Gantt chart for the execution
	of this new experiments for Dask and Spark.
	The x-axis shows the time in second while the y-axis depicts the workers, with 
	one worker thread ID (Dask) and process ID (Spark) per line.
	We represent each task on the Gantt chart with its associated function's color.
	We again observe that Spark reports three times more processes than Dask threads.
	\HL{
		It is unclear what is the cause for reporting extra Spark processes.
		However, we found that both Dask and Spark use the same 
		amount of maximum concurrent processes or threads.
		Additionally, there was no significant difference in the average amount of 
		concurrent processes or threads between Dask and Spark.
	}
								
	Figure~{\ref{fig:dask_kmeans_gantt}} and Figure~{\ref{fig:spark_kmeans_gantt}}
	show that Dask has a significantly shorther makespan than Spark, {\SI{338}{\second}}
	and {\SI{409}{\second}} respectively.
	The main differences in makespan for Spark appear from a longer initial load of the data
	and a longer scheduling delay between the data load and the initial computation.
								
	For brevity we omitted the figures when varying the number of iterations as the results are similar to varying the number of workers.
														
	\subsection{BIDS App Example}
\begin{figure}[!h]
		\centering
		\includegraphics[clip,width=0.75\columnwidth]{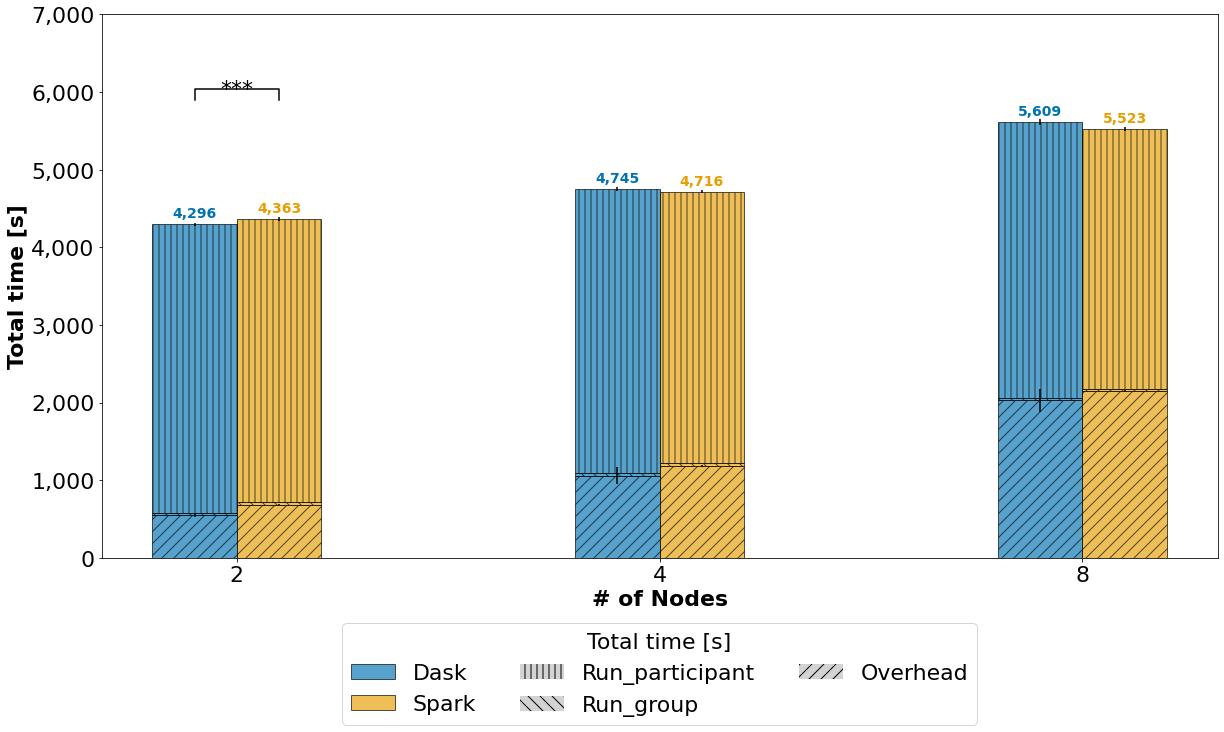}
		\caption{BIDS App Example: varying nodes}
		\label{fig:bids}
	\end{figure}

	Figure \ref{fig:bids} shows the total execution time for the \textit{BIDS App Example} application when varying the number of nodes.
	We observe that Dask has a slightly lower overhead, thank Spark, although not significant.
	However, the makespan for Dask is between 7.5\% to 14.5\% lower than for Spark.
	The main difference comes from an increased amount of stagger tasks for Spark, thus increasing the overhead of the application.
	Moreover, although minimal, a shorter cluster deployment time for Dask reinforces the difference in overhead time.
														
	\section{Discussion}
	\subsection{I/O bottleneck}
	\label{subsec:io-bottleneck}
	We used Lustre as shared file system as it is a de facto solution in HPC systems.
	The performance of our Lustre installation degrades when a large number of concurrent tasks attempt to perform I/O to the shared file system, as shown in Figures \ref{fig:increment_worker} and \ref{fig:histogram_worker}.
	In a previous study~\cite{8943502}, we observed nearly no improvement when increasing the number of nodes to process an application while using an NFS.
	In this paper, the performance scales when increasing the number of nodes, although not linearly due to an I/O bottleneck.
	The system was designed such that the filesystem has a similar speed to the network.
	Upgrading to more performant networking and storage components would reduce the bottleneck when increasing the number of workers.
								
	As in our previous study~{\cite{8943502}}, we observe that the I/O bottleneck has a lower impact on more compute-intensive applications, such as Kmeans and BIDS App Example.
	However, it significantly affects data-intensive applications, such as Increment and Histogram.

	These results suggest that increasing the parallelism of the file system and speed of the network is critical when scaling the compute resources to process a data-intensive application.
	Otherwise, the compute resource might be underutilized.
								
	\subsection{Tradeoff between overhead and data transfer}
	\label{subsec:overhead-tradeoff}
	Overall, except for Kmeans, the overhead for Dask is larger than for Spark.
	However, the total execution time for both engines is comparable.
	This almost exact compensation between the engine overhead and the data transfer time (see for example Figure \ref{fig:increment_worker}) is not surprising.
	When there is high contention on the bandwidth of the file system, and in the absence of trashing, desynchronizing the file transfers generally does not reduce the total data transfer time.
	To take an extreme example, the makespan of $n$ concurrent file transfer of size $F$ from a file system with a bandwidth $B$ would be $nF/B$,
	as the $n$ transfer would share the bandwidth $B$, which is equivalent to sequentially transferring the $n$ files.
	In a previous study~\cite{8943502}, we also observed this tradeoff between overhead and data transfer when the file system was saturated.
	Exploring strategies that take advantage of this tradeoff could be interesting to reduce the application makespan.
																		
	\subsection{Memory management}
	\HL{Our initial Kmeans results showed a significant difference between Dask and Spark.}
	The difference in performance \HL{came} from Spark worker restarting due to running out of memory; thus, Spark \HL{needed} to recompute several tasks.
	This is unexpected, given that Dask and Spark \HL{were} configured similarly and \HL{were} allocated the same amount of resources.
	We think those issues \HL{arose} from a combination of
	(1) the reserved memory for the JVM heap space,
	(2) the serialization of Python to Java kept in memory,
	(3) Spark using processes compared to Dask using threads, thus limiting shared memory consumption,
	and (4) different policies to persist data to disk when running out of memory.
								
	We optimized the memory usage in our Kmeans implementation for Dask and Spark
	and observed that the Spark workers no longer failed with this improved implementation.
	Nevertheless, Figure~{\ref{fig:dask_kmeans_gantt}} and Figure~{\ref{fig:spark_kmeans_gantt}} depicts that a substantial difference in makespan remained between Spark and Dask with this implementation,
	showing that the previously-based gap was not only due to worker failures.
								
	\subsection{Scheduling overhead}
	Overall, our results show that Dask has a slightly longer startup overhead than Spark.
	However, we found that this did not lead directly to the makespan difference.
	While both engines showed similar overhead for applications with low inter-worker communication,
	Spark had significantly more scheduling overhead than Dask for the Kmeans applications.
	Spark was considerably slower to schedule the first set of computations after reading the data on the workers.
	Moreover, Spark read all the data before starting computation while Dask was still reading data alongside doing computation.
	This difference in scheduling is beneficial for data-intensive applications as it reduces the risk
	of I/O bottleneck, as explained in Section~{\ref{subsec:overhead-tradeoff}}.
																		
	\subsection{Comparison to other studies}
	Overall, except for Kmeans, our experiments did not show substantial performance differences between Spark and Dask.
	In contradiction with the experiments of~\cite{Mehta:17}, we found that Dask had a larger overhead than Spark, especially when scaling the number of workers.
	However, this was compensated by the previously mentioned overhead and data transfer tradeoff.
	Since our focus is on \HL{neuroimaging} applications, we conclude that the performance of Dask and Spark is equivalent.
					
	\HL{
		In this study, we observed that Dask has a larger startup overhead than Spark.
		These results are consistent with a previous study~{\cite{Mehta:17}}.
		However, other studies found the opposite~{\cite{8588652}} or that the difference was negligible~{\cite{8943502}}.
	}
	Those apparent contractions most likely come from differences in infrastructure and configuration of the engines.
																		
	Our results from the Histogram experiments show that the Dask Bags API was slower than Spark.
	The experiments from~\cite{10.1145/3225058.3225128} also found that Dask Bags performed slower than Spark.
	We conclude that when using Dask, the other Dask APIs should be favored unless Dask Bags are strictly required.
																		
	Our results showed that, for data-intensive applications, increasing the number of workers should only be done if the bandwidth of the file system is not saturated.
	Two other studies support this~\cite{8943502, 8588652}.
	Therefore, it is critical to ensure enough bandwidth to the file system before scaling the compute resources for data-intensive applications.
	Otherwise, those extra resources might be underused.
	
	\HL{
		Another study~{\cite{9355308}} developed a parametrized benchmark framework
		to evaluate the performance of different engines, including Spark and Dask.
		On the one hand, the framework reduces the implementation effort and helps with
		comparing frameworks fairly.
		On the other hand, it is hard to characterize neuroimaging realistically
		applications using their framework since the runtime of these applications can vary
		greatly depending on the subject process, and the framework lacks kernels with
		a variable amount of compute-intensity.
		Due to these limitations, we decided to implement the benchmarked applications
		directly into Spark and Dask.
	}
					
	\subsection{Recommendations}
	\HL{
		The results from our experiments show that the performance difference between
		Dask and Spark are minimal, in general. 
		The main differences we observed are qualitative.
		We found that Dask integration with the scientific Python ecosystem made it
		easier to implement neuroimaging applications.
		Spark configuration required to set multiple environment variables.
		In comparison, Dask is ready-to-go after an installation through PIP, the Python package manager.
		The Dask dashboard provides more information than Spark's, making
		the debugging of Dask applications easier and faster.
		Therefore, for the neuroimaging field, we recommend the use of Dask over Spark.
		However, in other fields with non-Python-based ecosystem, both Dask and Spark
		may be equal candidates as a Big Data engine.
	}
	
	\HL{
		While the engines showed similar performance, they both come with their caveats that
		can drastically affect their performance.
		For Dask, the user should be aware of the Python GIL limitations.
		For Python code, multiples worker processes should be used while for code releasing the
		GIL, a balance between processes and thread is recommended.
		For Spark, we experienced severe performance issues due to workers running out of memory.
		Users should carefully consider their Spark worker configurations to prevent similar issues.
	}
												
	\subsection{Limitations}
	To make a fair comparison between Spark and Dask, we set both engine configurations
	to be equivalent to the best of our capabilities, which comes with some limitations.
	Firstly, both engines use a shared implementation for the algorithms.
	This is done to keep the theoretical runtime equivalent; however, the chosen
	implementation might not be optimal for either of the engines.
	Secondly, we decided to use Lustre as a file system for both Spark and Dask.
	Spark could benefit from using HDFS as a file system, but HDFS is uncommon in HPC.
	\HL{Finally, our Lustre filesystem suffers from an I/O bottleneck coming from 
	either the network or disk bandwidth.}
	As explained in section~{\ref{subsec:overhead-tradeoff}}, the data transfer 
	time is tightly correlated to the overhead of the application.
	However, while reducing this bottleneck should improve the performance of 
	both engines similarly, it can be hard to predict the actual behavior.
	
	\HL{
		While both Dask and Spark support various high-performance network strategies
		such as UCX, we focused on the standpoint of neuroscientists that would 
		commonly use the frameworks mostly out of the box.
		With higher network throughput, the overall performance of the engines could differ.
		Further work would be required to compare Dask and Spark using different high-performance
		network protocols.
	}
				
	\HL{
		Dask and Spark both support GPUs; however, we did not use applications that
		leverage them in our work.
		Although several neuroimaging applications use GPUs, only a few preprocessing pipelines use them.
		Therefore, we omitted GPUs from this study because preprocessing comprises the majority of computing time for neuroimaging
		workflow.
		Moreover, the few preprocessing applications using GPUs are not commonly used
		due to their recent development.
	}
	
	\HL{
		Due to the limited amount of benchmarks available to evaluate the performance
		of neuroimaging applications, we use our own collection of applications.
		Further efforts would be needed to create a credible and in-depth performance
		benchmark specific to the neuroimaging field.
	}
												
	\section{Conclusion}
	We presented a detailed comparison of two Big Data engines, Apache Spark and Dask.
	We studied the engines using three synthetic \HL{neuroimaging} applications and 
	a neuroimaging pipeline \HL{to calculate brain volume.}
	Overall, our results showed no substantial difference between the engines.
	Although Spark overhead was generally lower than Dask, it was significantly higher when applications incur inter-worker communication.
	For one application, our results showed that Spark used more memory, causing issues with its workers resulting in a significantly faster runtime for Dask.
	However, using an infrastructure offering more memory per worker would most likely solve this issue.
																						
	These results suggest that future research should focus on strategies to:
	\begin{itemize}
		\item Limit the impact of data transfers for data-intensive applications
		\item Reduce the footprint of application memory
		\item Manage the memory of workers more efficiently
	\end{itemize}
	
	\section*{Acknowledgment}
	Mathieu Dugr\'e was partially funded by a Alexander Graham Bell scholarship from the Natural
	Sciences and Engineering Research Council of Canada and a Master's (B1X) Research Scholarships.
	The Canada Research Chairs program partially funded this work.
	The Canada Foundation for Innovation funded the HPC cluster.
	
	\section*{Data Availability}
	\subsection*{BigBrain}
	The data that support the findings of this study are available from the corresponding author, M. Dugr\'e, upon reasonable request.
	A lower resolution of \SI{40}{\micro\metre} is available publicly by FTP at: 
	\newline\href{https://ftp.bigbrainproject.org/bigbrain-ftp/BigBrainRelease.2015/3D_Blocks/40um/nii/}{https://ftp.bigbrainproject.org/bigbrain-ftp/BigBrainRelease.2015/3D\_Blocks/40um/nii}
	
	\subsection*{CoRR}
	The data that support the findings of this study are openly available in DataLad at:
	\newline\href{https://datasets.datalad.org/?dir=/corr/RawDataBIDS}{https://datasets.datalad.org/?dir=/corr/RawDataBIDS}
	
	\subsection*{Code}
	The code that support the findings of this study is openly available in GitHub at:
	\newline\href{https://github.com/big-data-lab-team/paper-big-data-engines/tree/CPE}{https://github.com/big-data-lab-team/paper-big-data-engines/tree/CPE}
	
	\section*{Conflict of Interests}
	The authors declare no conflict of interests.
	
	\bibliography{paper}

\begin{thebibliography}{10}
\providecommand \doibase [0]{http://dx.doi.org/}%

\bibitem{ALFAROALMAGRO2018400}
Alfaro-Almagro F, Jenkinson M, Bangerter NK, et al. Image processing and
  Quality Control for the first 10,000 brain imaging datasets from UK Biobank.
  {\it NeuroImage} 2018\string; 166\string: 400-424.
\newblock \href {\doibase https://doi.org/10.1016/j.neuroimage.2017.10.034}
  {doi: https://doi.org/10.1016/j.neuroimage.2017.10.034}

\bibitem{van2014human}
Van~Horn JD, Toga AW. Human neuroimaging as a "{B}ig {D}ata" science. {\it
  Brain imaging and behavior} 2014\string; 8(2)\string: 323--331.

\bibitem{ConpPortal}
CONP . Canadian Open Neuroscience Platform (CONP) Portal.
  \href{https://portal.conp.ca}{https://portal.conp.ca}; .

\bibitem{Nipype:11}
Gorgolewski K, Burns C, Madison C, et al. Nipype: A {F}lexible, {L}ightweight
  and {E}xtensible {N}euroimaging {D}ata {P}rocessing {F}ramework in {P}ython.
  {\it Frontiers in Neuroinformatics} 2011\string; 5\string: 13.
\newblock \href {\doibase 10.3389/fninf.2011.00013} {doi:
  10.3389/fninf.2011.00013}

\bibitem{8752675}
Hayot-Sasson V, Brown ST, Glatard T. Performance Evaluation of Big Data
  Processing Strategies for Neuroimaging. In: 2019 19th IEEE/ACM International
  Symposium on Cluster, Cloud and Grid Computing (CCGRID). ; 2019\string:
  449-458

\bibitem{Dask:15}
{R}ocklin M. {D}ask: {P}arallel {C}omputation with {B}locked algorithms and
  {T}ask {S}cheduling. In: {P}roceedings of the 14th {P}ython in {S}cience
  {C}onference. ; 2015\string: 126 - 132

\bibitem{Spark:16}
Zaharia M, Xin RS, Wendell P, et al. Apache {S}park: {A} {U}nified {E}ngine for
  {B}ig {D}ata {P}rocessing. {\it Commun. ACM} 2016\string; 59(11)\string:
  56--65.
\newblock \href {\doibase 10.1145/2934664} {doi: 10.1145/2934664}

\bibitem{hindman2011mesos}
Hindman B, Konwinski A, Zaharia M, et al. Mesos: A Platform for
  $\{$Fine-Grained$\}$ Resource Sharing in the Data Center. In: 8th USENIX
  Symposium on Networked Systems Design and Implementation (NSDI 11). ; 2011.

\bibitem{vavilapalli2013apache}
Vavilapalli VK, Murthy AC, Douglas C, et al. Apache {H}adoop {YARN}: {Y}et
  another resource negotiator. In: ACM. ; 2013\string: 5.

\bibitem{Braam2019TheLS}
Braam PJ. The Lustre Storage Architecture. {\it ArXiv} 2019\string;
  abs/1903.01955.

\bibitem{OpenSFS-lustre}
OpenSFS . Lustre® File System, Version 2.4 Released.
  \href{https://www.opensfs.org/press-releases/lustre-file-system-version-2-4-released/}{https://www.opensfs.org/press-releases/lustre-file-system-version-2-4-released/};
  .

\bibitem{8943502}
{Dugr\'e} M, {Hayot-Sasson} V, {Glatard} T. A Performance Comparison of Dask
  and Apache Spark for Data-Intensive Neuroimaging Pipelines. In: 2019 IEEE/ACM
  Workflows in Support of Large-Scale Science (WORKS). ; 2019\string: 40-49

\bibitem{8588652}
Chantzialexiou G, Luckow A, Jha S. Pilot-Streaming: A Stream Processing
  Framework for High-Performance Computing. In: 2018 IEEE 14th International
  Conference on e-Science (e-Science). ; 2018\string: 177-188

\bibitem{Mehta:17}
Mehta P, AlSayyad Y, Dorkenwald S, et al. Comparative evaluation of big-data
  systems on scientific image analytics workloads. {\it Proceedings of the VLDB
  Endowment} 2017\string; 10(11)\string: 1226-1237.
\newblock \href {\doibase 10.14778/3137628.3137634} {doi:
  10.14778/3137628.3137634}

\bibitem{10.1145/3225058.3225128}
Paraskevakos I, Luckow A, Khoshlessan M, et al. Task-Parallel Analysis of
  Molecular Dynamics Trajectories. In: ICPP 2018. Proceedings of the 47th
  International Conference on Parallel Processing. Association for Computing
  Machinery; 2018; New York, NY, USA

\bibitem{DaskBagDoc}
Dask Bag Documentation: Known Limitations.
  \url{https://docs.dask.org/en/latest/bag.html#known-limitations}; .
\newblock Accessed: 2022-08-17.

\bibitem{dean2008mapreduce}
Dean J, Ghemawat S. Map{R}educe: simplified data processing on large clusters.
  {\it Communications of the ACM} 2008\string; 51(1)\string: 107--113.

\bibitem{RDD}
Zaharia M, Chowdhury M, Das T, et al. Resilient {D}istributed {D}atasets: {A}
  {F}ault-tolerant {A}bstraction for {I}n-memory {C}luster {C}omputing. In:
  NSDI'12. Proceedings of the 9th USENIX Conference on Networked Systems Design
  and Implementation. USENIX Association; 2012; Berkeley, CA, USA\string: 2--2.

\bibitem{DaskSLURMDoc}
Dask SLURM Cluster Documentation.
  \url{https://jobqueue.dask.org/en/latest/generated/dask_jobqueue.SLURMCluster.html#dask-jobqueue-slurmcluster};
  .
\newblock Accessed: 2022-08-23.

\bibitem{Amunts:13}
Amunts K, Lepage C, Borgeat L, et al. Big{B}rain: An {U}ltrahigh-{R}esolution
  3{D} {H}uman {B}rain {M}odel. {\it Science} 2013\string; 340(6139)\string:
  1472--1475.
\newblock \href {\doibase 10.1126/science.1235381} {doi:
  10.1126/science.1235381}

\bibitem{sam}
SAM Library. \url{https://github.com/big-data-lab-team/sam}; .
\newblock Accessed: 2022-09-12.

\bibitem{zuo2014open}
Zuo XN, Anderson JS, Bellec P, et al. An open science resource for establishing
  reliability and reproducibility in functional connectomics. {\it Scientific
  data} 2014\string; 1\string: 140049.

\bibitem{DataladDataset}
Datalad dataset. \url{https://datasets.datalad.org/?dir=/corr/RawDataBIDS}; .
\newblock Accessed: 2022-09-12.

\bibitem{gorgolewski2017bids}
Gorgolewski KJ, Alfaro-Almagro F, Auer T, et al. {BIDS} apps: {I}mproving ease
  of use, accessibility, and reproducibility of neuroimaging data analysis
  methods. {\it {PLoS} computational biology} 2017\string; 13(3)\string:
  e1005209.

\bibitem{9355308}
Slaughter E, Wu W, Fu Y, et al. Task Bench: A Parameterized Benchmark for
  Evaluating Parallel Runtime Performance. In: SC20: International Conference
  for High Performance Computing, Networking, Storage and Analysis. ;
  2020\string: 1-15

\end{thebibliography}
\end{document}